\documentclass[journal, 10pt]{IEEEtran}

\usepackage{multirow}
\usepackage[font=scriptsize,caption=false,labelsep=space]{subfig}
\usepackage{mathrsfs}
\usepackage{graphicx,float,wrapfig,epstopdf,amsmath}
\usepackage[]{algorithmicx}
\usepackage{algpseudocode,algorithm}
\usepackage{balance}
\usepackage{mathtools,lipsum}
\usepackage{amsmath}
\usepackage{amssymb}
\usepackage{amsthm}
\usepackage{graphicx}
\usepackage{epstopdf}
\usepackage{times}
\usepackage{textcomp,cite}

\usepackage{url}
\usepackage{hyperref}

\usepackage{multirow}
\usepackage{threeparttable}
\graphicspath{{./figures/}}

\usepackage[table]{xcolor}
\definecolor{intnull}{RGB}{213,229,255}
\definecolor{inteins}{RGB}{128,179,255}
\definecolor{color1}{RGB}{199,209,232}
\definecolor{color2}{RGB}{230,231,233}

\begin{document}

	\title{The Rise of Intelligent Reflecting Surfaces in \\ Integrated Sensing and Communications Paradigms}

	\author{\IEEEauthorblockN{Ahmet M. Elbir, \textit{Senior Member, IEEE,} Kumar Vijay Mishra, \textit{Senior Member, IEEE,} M.~R.~Bhavani~Shankar, \textit{Senior Member, IEEE,} and Symeon Chatzinotas, \textit{Fellow, IEEE}}
		\thanks{This work was supported in part by the  ERC project AGNOSTIC and FNR project SPRINGER.}
		\thanks{A. M. Elbir is with 	the Interdisciplinary Centre for Security, Reliability and Trust (SnT)  at the University of Luxembourg, Luxembourg and with the Department of Electrical and Electronics Engineering, Duzce University, Duzce, Turkey (e-mail: ahmetmelbir@gmail.com).} 
		\thanks{K. V. Mishra is with the United States Army Research Laboratory, Adelphi, MD 20783 USA, and with the SnT at the University of Luxembourg, Luxembourg (e-mail: kumarvijay-mishra@uiowa.edu). }
		\thanks{M. R. B. Sankar and S. Chatzinotas are with the SnT at the University of Luxembourg, Luxembourg (email: \{bhavani.shankar, symeon.chatzinotas\}@uni.lu). }	
	}
	\maketitle
	
	\begin{abstract}
		The intelligent reflecting surface (IRS) alters the behavior of wireless media and, consequently, has potential to improve the performance and reliability of wireless systems such as communications and radar remote sensing. Recently, integrated sensing and communications (ISAC) has been widely studied as a means to efficiently utilize spectrum and thereby save cost and power. This article investigates the role of IRS in the future ISAC paradigms. While there is a rich heritage of recent research into IRS-assisted communications, the IRS-assisted radars and ISAC remain relatively unexamined. We discuss the putative advantages of IRS deployment, such as coverage extension, interference suppression, and enhanced parameter estimation, for both communications and radar. We introduce possible IRS-assisted ISAC scenarios with common and dedicated surfaces. The article provides an overview of related signal processing techniques and the design challenges, such as wireless channel acquisition, waveform design, and security.
	\end{abstract}


	\section{Introduction}
	\label{sec:Introduciton}
	Radar and communications systems have witnessed tremendous progress for several decades while exclusively operating in different frequency bands to minimize the interference to each other~\cite{jrc_overview_TCOM}. Modern radar systems consume a large portion of the spectrum - from very-high-frequency (VHF) to Terahertz (THz) - in various applications, such as over-the-horizon, air surveillance, meteorological, military, and automotive radars. Similarly, communications systems have moved forward from ultra-high-frequency (UHF) to millimeter-wave (mm-Wave) in response to the demand for providing new services and accommodating massive number of users with high data rate requirements for the applications, such as high-definition video transmission, intra- and inter-vehicular messaging, machine-to-machine communications, and internet-of-things  architectures. The mm-Wave band has also attracted various radar applications, such as automotive radar ($24$-$80$ GHz), indoor localization ($260$ GHz), and cloud observation ($95$ GHz)~\cite{jrc_overview_TCOM}. Furthermore, the THz bands have gained much interest recently for short-range communications and radar because it offers ultra-wide bandwidth and extremely high angular/range resolution~\cite{elbir2021JointRadarComm}. 
	
	\textcolor{black}{The aforementioned advances and the prior fragmented allocation of frequency bands lead to the inefficient use of the spectrum.} The carrier aggregation or spectrum stitching techniques employed to address spectral congestion in communications requires specialization of the system components. In radar, such techniques suffer from lack of phase synchronization. It is, therefore, essential to develop strategies to simultaneously and opportunistically operate in the identical spectral bands in a mutually beneficial manner. As a result, there has been a substantial interest on jointly accessing the spectrum in an \textit{integrated sensing and communications} (ISAC) set-up~\cite{jrc_overview_TCOM}. Broadly, ISAC designs follow two directions: radar-communications coexistence (RCC) and dual-functional radar-communications (DFRC)~\cite{elbir2021JointRadarComm}. While the former involves efficient interference and resource management techniques so that both systems operate without unduly interfering each other, the latter focuses on designing ISAC systems to simultaneously perform communications and radar tasks. The existing mm-Wave communications protocols/waveforms are also attracting much investigation in this context. For example, the $60$ GHz IEEE 802.11ad standard wi-fi protocol has been proposed for communications-aided vehicular sensing~\cite{jrc_overview_TCOM}.

	
	In communications networks, the demand for higher frequency bands has led to the deployment of massive number of antenna arrays -  usually in a multiple-input
	multiple-output (MIMO) configuration - to cope with the high path loss arising from attenuation, scattering, reflection, and refraction. In addition, hybrid analog and digital beamforming is employed to reduce the number of radio frequency (RF) chains and hardware complexity.  To further reduce the hardware and power consumption, low-cost densely-packed planar arrays in the form of \textit{intelligent reflective 	surfaces} (IRSs) --- the literature also suggests the usage of other terms such as reconfigurable intelligent surfaces and large intelligent surfaces --- are envisaged in the sixth generation (6G) systems as a promising solution~\cite{irs_Overview_Pan2022Aug}.

	\begin{figure*}[t]
		\centering
		{\includegraphics[draft=false,width=.99\textwidth]{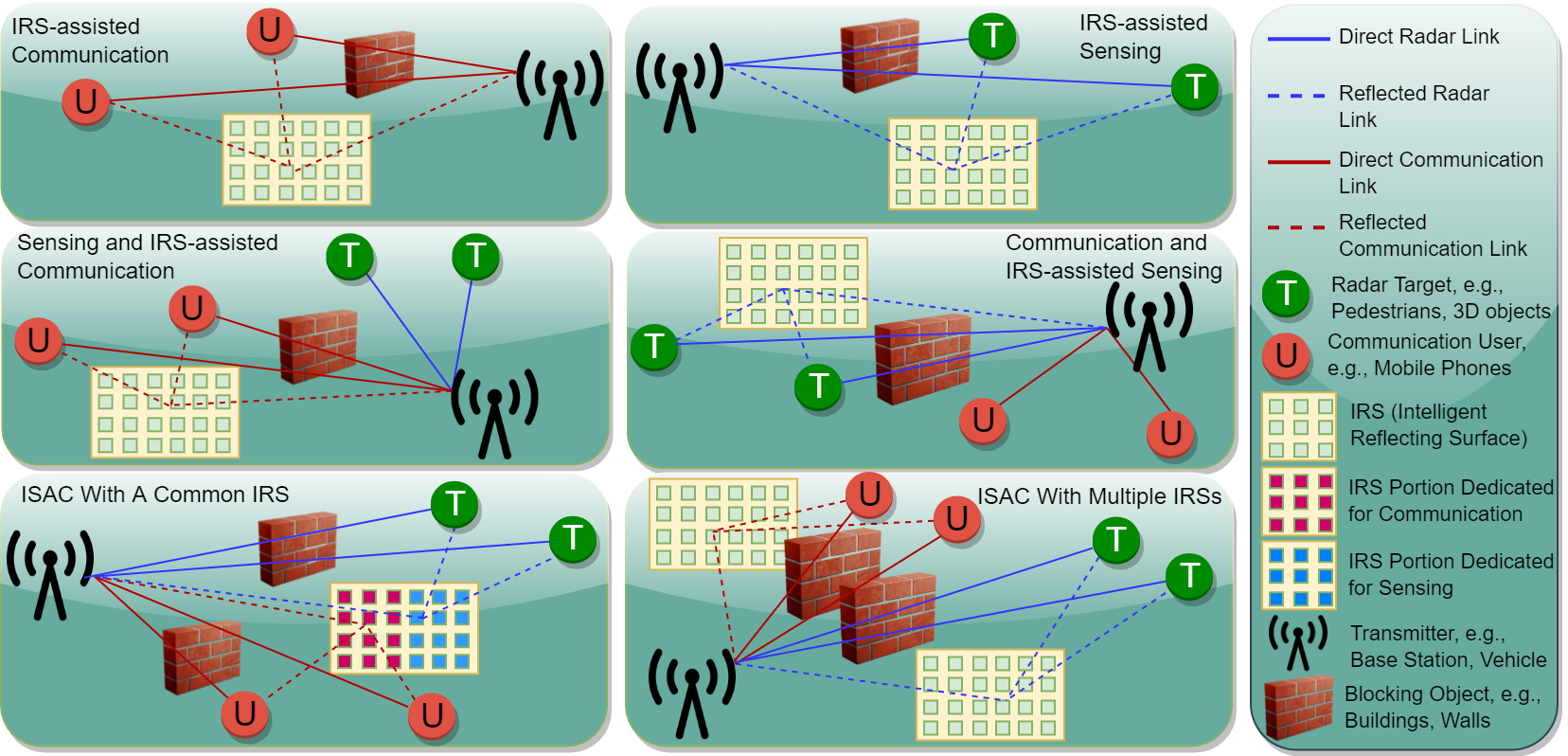} }
		\caption{Typical use cases of IRS in communications (top left); sensing (top right); sensing and IRS-assisted communications (middle left); communications and IRS-assisted sensing (middle right); ISAC with a common IRS (bottom left); and ISAC with multiple IRSs (bottom right).
		} \vspace{-10pt}
		\label{figIRS_ISAC}
	\end{figure*}

	\begin{table*}
		\caption{State-of-the-art in IRS-assisted ISAC
		}
		\label{tableSummary}
		\centering
		\begin{tabular}{p{0.13\textwidth}p{0.21\textwidth}p{0.32\textwidth}p{0.24\textwidth}}
			\hline 
			\hline
			\cellcolor{color2}\bf IRS Deployment &\bf Application\cellcolor{color1} & \cellcolor{color2}\bf Signal Processing Techniques
			& \cellcolor{color2}\bf Performance Metric / Constraint\\
			\hline
			\cellcolor{color2}	Communications \cite{irs_Overview_Pan2022Aug}
			& \cellcolor{color1}  Active/passive beamforming
			& \cellcolor{color2}  Alternating optimization and SDR
			& \cellcolor{color1}Total transmit power\\
			\hline
			\cellcolor{color2}	Sensing \cite{irs_metasensing}
			& \cellcolor{color1} 3D object detection
			& \cellcolor{color2} Deep reinforcement learning
			& \cellcolor{color1} Cross-entropy loss of the beampattern\\
			\hline
			\cellcolor{color2}	Sensing \cite{irs_radarDetectionBuzziTSP}
			& \cellcolor{color1}  Far-field and near-field target detection
			& \cellcolor{color2} Forward-backward IRS optimization via alternating maximization
			& \cellcolor{color1} SNR and probability of detection\\
			\hline
			\cellcolor{color2}	Sensing \cite{irs_radar_activePassiveBeamforming}
			& \cellcolor{color1} Multi-IRS optimization for detection in cluttered environment
			& \cellcolor{color2}Alternating optimization of radar beamformer and IRS phases via SDR
			& \cellcolor{color1}Minimum target illumination power\\
			\hline
			\cellcolor{color2}	ISAC \cite{irs_jrc_IRSpartitioning}
			& \cellcolor{color1} Partitioning IRS elements for localization and communication
			& \cellcolor{color2} Codebook-based design for IRS phase-shifts  
			& \cellcolor{color1} MSE of the target direction cosine vector\\
			\hline
			\cellcolor{color2}	ISAC \cite{irs_jrc_systemsJournal}
			& \cellcolor{color1} DFRC (single user, single target) 
			& \cellcolor{color2} SDR and Bisection search for transmit beamforming; majority-minimization for IRS design
			& \cellcolor{color1}SNR at the radar receiver  \\
			\hline
			\cellcolor{color2}	ISAC \cite{irs_jrc_txrxbeamforming}
			& \cellcolor{color1} DFRC (single user, multi-target)
			& \cellcolor{color2}Alternating optimization of transmit beamformer and the IRS phase shifts via SDR
			& \cellcolor{color1}IRS beampattern gain with communications SNR constraint\\
			
			\hline
			\cellcolor{color2}\textcolor{black}{ISAC \cite{jrc_irs_waveform_Liu2022May}}
			& \cellcolor{color1} \textcolor{black}{DFRC (multi-user, single target) with multiple clutters }
			& \cellcolor{color2} \textcolor{black}{ADMM and majorization-minimization for joint waveform, receive filter and  beamformer design}
			& \cellcolor{color1} \textcolor{black}{Radar SINR with quality-of-service constraints} \\
			\hline
			\cellcolor{color2}	ISAC \cite{irs_jrc_multipleIRS}
			& \cellcolor{color1} Wideband DFRC with multiple IRSs (multi-user, single target)
			& \cellcolor{color2} Dinkelbach's method for transmit beamforming; ADMM for IRS design
			& \cellcolor{color1}Average radar SINR plus  communications SINR \\
			\hline
			\cellcolor{color2}	ISAC \cite{irs_jrc_interferenceMitigation_Wang}
			& \cellcolor{color1} DFRC   (multi-user, multi-target)
			& \cellcolor{color2} Manifold optimization for transmit beamforming; SDR for IRS design
			& \cellcolor{color1}Multi-user interference with radar beampattern constraint  \\
			\hline
			\cellcolor{color2}	ISAC \cite{irs_jrc_DiscretePhaseShift}
			& \cellcolor{color1} DFRC  with quantized IRS phases (multi-user, multi-target)
			& \cellcolor{color2}Manifold optimization for transmit beamforming; successive optimization for IRS design
			& \cellcolor{color1} communications MSE and the  CRLB of the target DoAs  \\
			\hline
			\cellcolor{color2}	\textcolor{black}{ISAC \cite{irs_jrc_SecrecyRateOptMishra}}
			& \cellcolor{color1} \textcolor{black}{Secure DFRC in the presence of eavesdropping targets}
			& \cellcolor{color2}\textcolor{black}{Stochastic gradient descent to find precoders for information and artificial noise}
			& \cellcolor{color1} \textcolor{black}{Secrecy rate and MIMO radar SINR}\\
			\hline
			\cellcolor{color2}	\textcolor{black}{ISAC \cite{irs_jrc_clutter_Liao2022Aug}}
			& \cellcolor{color1} \textcolor{black}{DFRC (multi-user, multi-target) with multiple clutters     }
			& \cellcolor{color2}\textcolor{black}{Alternating optimization with semi-definite relaxation  }
			& \cellcolor{color1}\textcolor{black}{ Minimum sensing beampattern gain with SINR constraint   }\\
			\hline
			\hline
		\end{tabular}
	\end{table*}
	

	An IRS is composed of a large periodic array of subwavelength scattering meta-material elements, which reflect the incoming signal by introducing a pre-determined phase shift, and form an electrically-thin two-dimensional (2D) surface~\cite{irs_Overview_Pan2022Aug}. In wireless network applications, the phase shift corresponding to each element is reconfigured in real-time via external signals by the base station (BS) through a backhaul control link to manipulate the direction of the incoming signal from the BS toward the users. This specific usage of IRS improves the received signal energy at the distant users and expands the coverage of the BS. \textcolor{black}{The IRS-assisted systems enable analog beamforming by simply configuring these phase shifts so that the BS only performs digital beamforming with smaller number of antennas, thereby, reducing the hardware cost and improving energy efficiency. Nevertheless, the backhaul control of the IRS entails a fair contribution to the overall power consumption.}

	Lately, the benefits of IRS have been extensively analyzed in communications to enhance energy-efficiency, improve channel statistics/estimates, and network coverage~\cite{irs_Overview_Pan2022Aug}. However, corresponding studies for IRS-assisted radar and ISAC are relatively recent and fewer (see Table~\ref{tableSummary} for the summary of these recent works).	This article, therefore, presents various IRS-assisted radar and ISAC models (see Fig.~\ref{figIRS_ISAC}) that 
	utilize the IRS to provide flexibility for dynamic and accurate beamforming resulting in advantages such as improved spectral efficiency, energy-efficiency, coverage, parameter estimation, and interference suppression. We introduce possible IRS-assisted ISAC scenarios with common and dedicated surfaces.	Finally, we extensively discuss the key design challenges  and related signal processing techniques for IRS-assisted ISAC such as channel equalization, waveform design and physical layer security.

	\begin{figure*}[t]
		\centering
		{\includegraphics[draft=false,width=.98\textwidth]{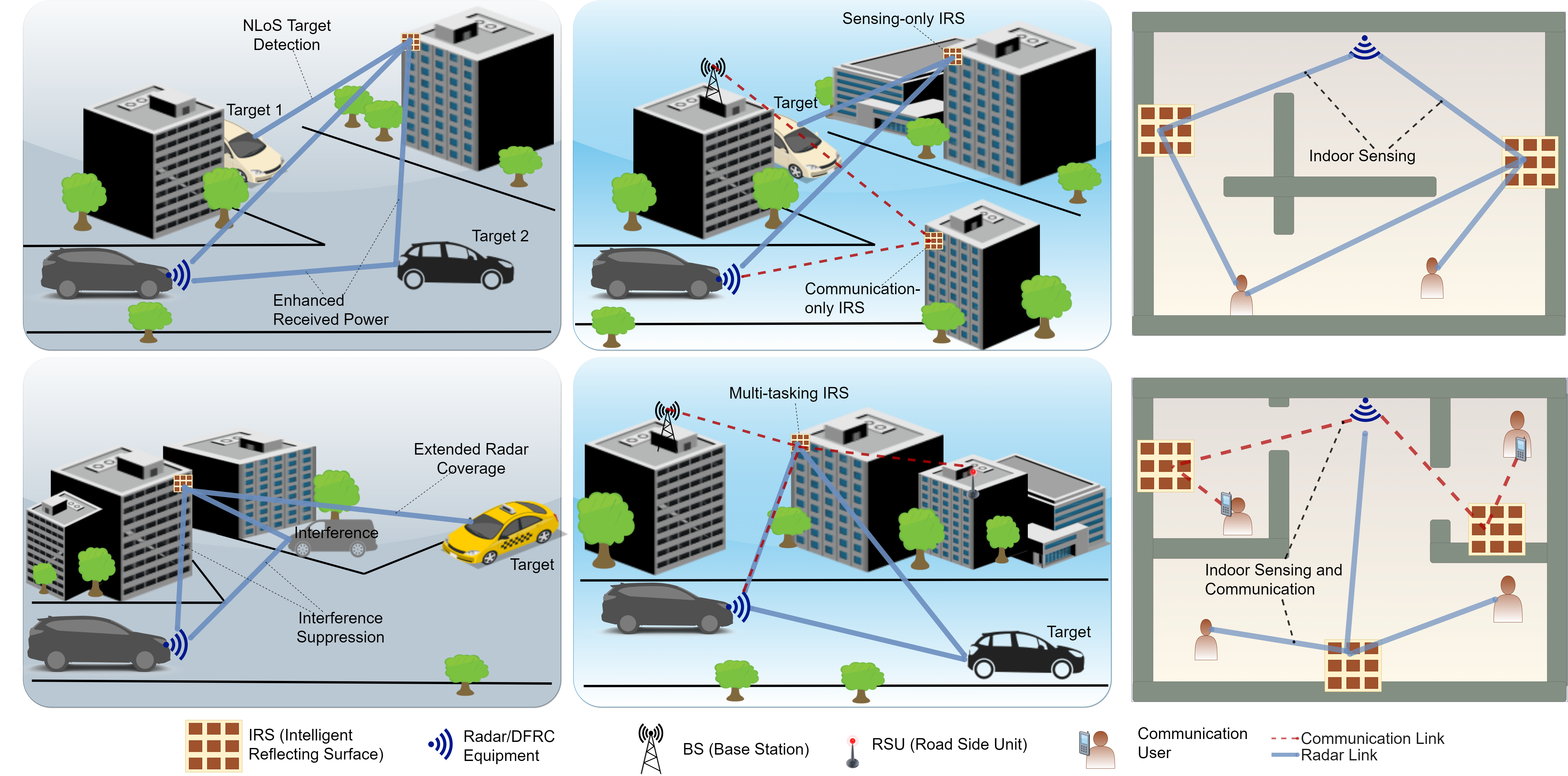} }
		\caption{Applications of IRS-aided wireless systems. The IRS-aided radars (top and bottom left) offer NLoS target detection, enhance received signal power, suppress interference, and extend the radar coverage. The IRS-aided ISAC (top and bottom middle) may facilitate individual communications and sensing purposes or both. The indoor IRS-aided systems (top and bottom right) have different channel dynamics and have typically been used for NLoS sensing and communications.} \vspace{-10pt}
		\label{figIRSRadISAC}
	\end{figure*}
	
	\section{IRS-Assisted Stand-Alone Systems}
	Given a large body of work on IRS-assisted communications (see, e.g. \cite{irs_Overview_Pan2022Aug} and references therein), we only briefly review the IRS-assisted communications before introducing the use cases of IRS in radar and ISAC.
	\subsection{IRS-Assisted Communications} 
	The IRS-aided communications generally deploy the surface between the BS and mobile users to improve the received signal power. 
	In some applications, it is also helpful to focus the incoming signal from the BS to the distant users using IRS, especially when the users are in the non-line-of-sight (NLoS) region. However, the IRS-assisted extension of the network coverage is limited by the fact that the surface is composed of passive elements; this is the key difference between IRS and active amplify-and-forward (AF) relays~\cite{irs_Overview_Pan2022Aug}.  Nevertheless, IRS is effective for indoor applications, wherein it is mounted on the walls of buildings to provide an additional link between the BS and indoor mobile users that may be inaccessible via LoS paths with conventional network structure~\cite{jrc_overview_TCOM}.

	\subsection{IRS-Assisted Radar}
	Unlike communications, backscatter signal carries information about the unknown channel or targets in radar applications. The IRS deployment is essentially aimed at focusing the target backscatter to the radar receiver (which may or may not be co-located with the radar transmitter). While IRS deployment in radar brings some of the similar advantages as in the communications e.g., extension of coverage and interference suppression, the signal processing at the receiver is different and aimed at extracting the unknown radar channel. \textcolor{black}{Hence, compared to conventional radar, IRS-assisted radar introduces the signal processing challenges such as  transmit waveform design and handling clutter as well as multipath propagation through IRS.}	This distinction leads to following unique applications.

	\begin{figure*}[t]
		\centering
		{\includegraphics[draft=false,width=\textwidth]{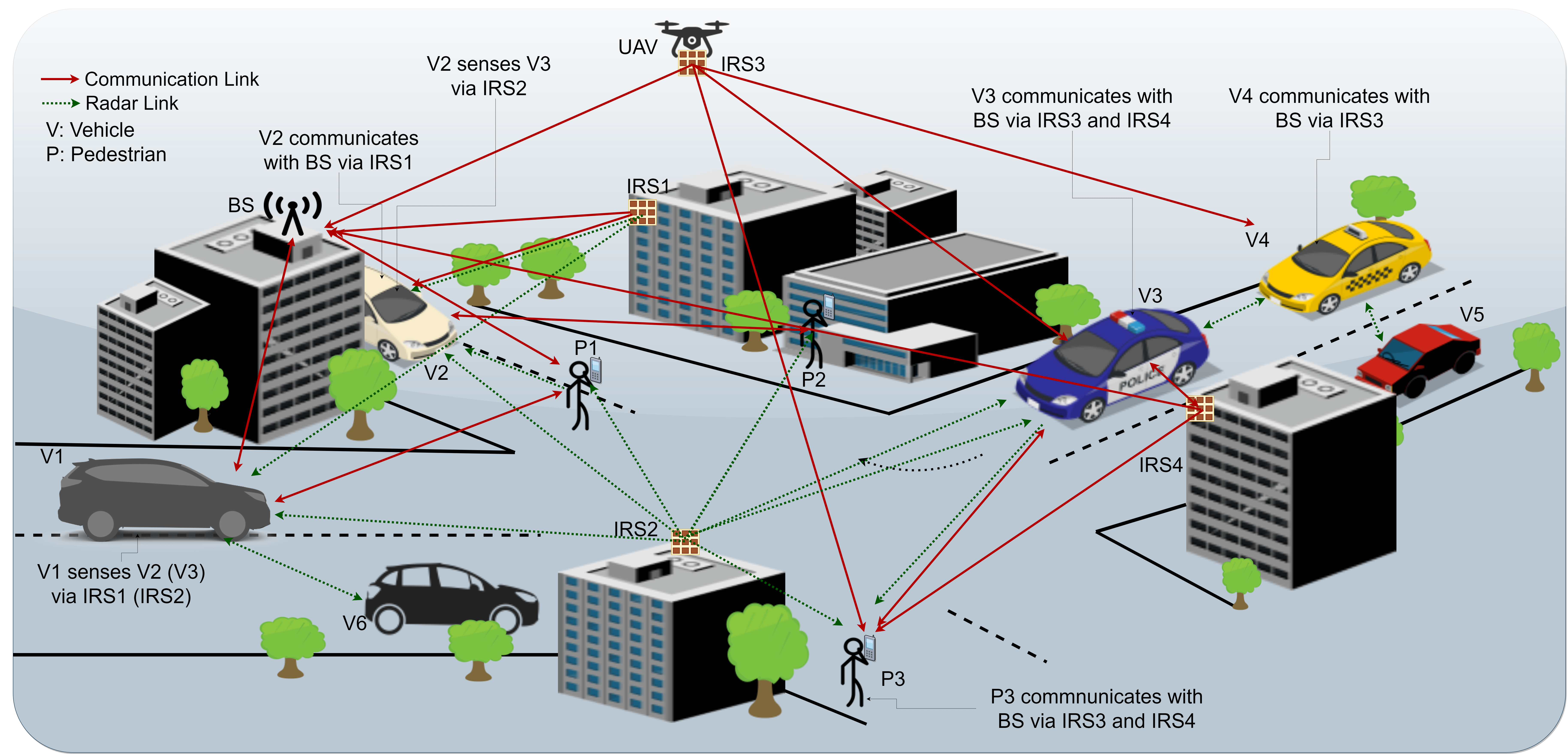} }
		\caption{\color{black}IRS-assisted RCC for various scenarios. The BS serves the communications users P3 and V3 via multiple surfaces. Vehicles V1, V2, V3 and V4 detect and track the targets through direct (V1-V6, V3-V4 and V4-V5) and IRS-assisted (V1-IRS1-V2, V1-IRS2-V3 and V2-IRS2-V3) paths; and jointly perform communications tasks via both direct (V1-P1, V2-P2 and V3-P3) and IRS-assisted (V2-IRS-1-BS, V3-IRS3-BS and V4-IRS3-BS) paths, respectively.
		} \vspace{-10pt}
		\label{figIRS_ISAC2}
	\end{figure*}

	\subsubsection{Detection of NLoS targets}
	The NLoS radar applications are mostly categorized as looking through a diffuser (e.g., through-the-wall radar) and looking around corner. While there has been an extensive research on the former, the latter has been recently paid a great attention driven by the developments of autonomous driving and smart city concepts~\cite{irs_radar_activePassiveBeamforming}. In such urban environments, the detection of the NLoS targets is an interesting and very challenging issue for both military and civilian applications, wherein the  moving targets such as vehicles, pedestrians and unmanned aerial vehicles (UAVs) may fall into the shadow region of the radar (e.g., behind the  buildings). In such scenarios, IRS-assisted techniques can be employed to improve the signal-to-noise-ratio (SNR) by leveraging the IRS-reflected radar link. For instance, as illustrated in Fig.~\ref{figIRSRadISAC}, the radar equipment mounted on the vehicle has no LoS to the vehicle behind the building (Target 1). Instead, the radar senses the target with the aid of IRS which constructs a reflected radar link. The IRS-aided system decides the LoS/NLoS path based on signal power analysis~\cite{irs_radar_activePassiveBeamforming}.  \textcolor{black}{When the radar receives reflected signals from the target and IRS simultaneously, these signals should should be resolved in range/angle/Doppler domain. Then, the IRS-assisted NLoS path can be distinguished due to received power after detecting them above a certain threshold.     }
	
	The joint active (passive) beamforming design at the radar (the IRS)  constitutes a complex optimization problem, for which alternating optimization techniques can be employed~\cite{irs_radar_activePassiveBeamforming,irs_radarDetectionBuzziTSP}. Such algorithms usually employ the performance metrics, such as the SNR~\cite{irs_radarDetectionBuzziTSP}, mean-squared-error (MSE) of the radar cross-section and the minimum target illumination power~\cite{irs_radar_activePassiveBeamforming}. Multiple IRS deployments may also be employed to facilitate indirect path sensing in other directions~\cite{irs_radar_activePassiveBeamforming,irs_jrc_multipleIRS}. For instance in~\cite{irs_radar_activePassiveBeamforming}, the phase shifts of two IRSs are jointly optimized to sense the multiple targets in a cluttered environment. 
	
	\subsubsection{Enhanced SNR and Extending Coverage}
	With the aid of IRS-assisted radar link, the radar detects/tracks the NLoS targets with enhanced SNR (Fig.~\ref{figIRSRadISAC}). Even for LoS targets, it is possible to enhance the received signal power by utilizing the IRS-reflected signals. 
	In~\cite{irs_radarDetectionBuzziTSP}, an IRS-assisted MIMO radar operating at $3$ GHz 
	transmits probing signals with narrow beams and receives the echoes reflected from both target and IRS (whose phases are optimized and controlled accordingly by the radar via backhaul link). 
	
	\textcolor{black}{The enhanced SNR also allows the extension of radar coverage (Fig.~\ref{figIRSRadISAC}), which  is particularly helpful in long range radar (LRR) applications, wherein more transmit power is required to detect the distant targets. In~\cite{irs_radarDetectionBuzziTSP}, it is shown that using a single or multiple IRSs, up to $10$ dB SNR gain can be achieved for closely positioned IRSs. The SNR gain fades away as the distance increases while still maintaining higher gain compared to no-IRS radar case.}


	\subsubsection{Interference suppression}
	The passive IRS beamforming is useful to suppress the interference from coexisting emitters while maintaining a satisfactory radar performance~\cite{irs_jrc_DiscretePhaseShift}.  \textcolor{black}{Although the automotive radars have highly directive beams, which can suppress the signals from other directions, the interference signal can be significant, for which IRS-reflected signals can be employed. As shown in Fig.~\ref{figIRSRadISAC}, direct and IRS-reflected signals from the interference source may add destructively to cancel out the interference.} 	

	\subsubsection{Indoor sensing}
	The NLoS targets are very common in indoor applications (Fig.~\ref{figIRSRadISAC}), where targets may be occluded by walls. Here, without any sophisticated and heavily noise-ridden through-the-wall imaging, IRS-aided radar detects and localizes the targets.
	In order to improve the sensing performance of the radar, an IRS-assisted RF sensing method was devised  in~\cite{irs_metasensing}. Specifically, a pair of single-antenna transceivers operating at $3.2$ GHz are used to transmit and receive signals from the IRS to sense a 3D object. To this end, the beampattern of the IRS elements is designed by configuring the IRS elements via Markov decision process and a deep reinforcement learning technique is employed to optimize the cross-entropy loss of the sensing accuracy. Here, a large number of IRS elements improve the sensing accuracy of 3D targets up to twice the performance of non-IRS system. Further, multiple IRS offers significant improvement in resolution and mitigates occlusions, in a manner similar to the distributed radar system, but without the active transmission~\cite{irs_jrc_multipleIRS}. In addition, the time-difference-of-arrivals (TDoAs) of direct and IRS-reflected paths may also be utilized to improve localization performance~\cite{irs_radarDetectionBuzziTSP}. 
	
	\subsubsection{Aerial sensing}
	UAV-borne radars have become an effective remote sensing applications, such as for security and rescue in inclement environments and disaster sites. While UAVs have several advantages for both sensing and communications, such as 
	flexibility and ubiquitous connectivity, they have limited battery life (usually under $30$ minutes)~\cite{jrc_overview_TCOM}. In order to reduce the energy consumption, IRS-assisted schemes can be useful to extend the coverage of UAVs, thus reducing the extra movement of the UAV. As illustrated in Fig.~\ref{figIRS_ISAC2}, the UAV can also be equipped with an IRS, which operates as a relay  to improve the quality of service in a vehicular network. In such a scenario, the UAV-borne IRS can be utilized to extend the radar coverage in sensing applications. Another interesting application could be a smart city concept, wherein the IRSs placed on the building facades	are used to improve the received signal strength at the UAV radar flying well over the BS.


	%
	%

	%
	%
	%
	%
	%
	%
	%
	%
	%
	%

	\section{IRS-Aided ISAC}
	Various IRS deployment scenarios have been reported in the literature depending on \textcolor{black}{the common, coexistent, and individual functionalities.} Interestingly, each one of these gives rise to novel signal processing and optimization problems.

	\subsection{Common Deployment}
	Since the DFRC system employs a common transmitter, the common IRS usage for radar and communications tasks is more suitable.  As illustrated in Fig.~\ref{figIRS_ISAC2}, the IRS1 is commonly used by V2 for sensing V1 and communicating with BS. This IRS should distinguish the radar and communications signals to reflect them to users or targets accordingly. This is achieved via a protocol where the ISAC applications  operate  at  different time-/frequency-/spatial-/code- divisions to avoid mutual interference. \textcolor{black}{Such non-overlapping resource allocation techniques are easy to implement and have low complexity, the goal of ISAC design is to unify both sensing and communication task within a harmony. Recent works present overlapping resource management techniques to achieve fully-unified waveform design with increased degrees-of-freedom (DoF) while entailing high complexity~\cite{jrc_overview_TCOM}}.

	The joint design of the DFRC and IRS parameters is required in common deployment. Therefore, power-, hardware- and computation- efficient techniques are of great importance for joint design.   In	\cite{irs_jrc_IRSpartitioning}, an IRS-partitioning approach is proposed, wherein a portion of the IRS is dedicated for sensing tasks, while the remaining IRS elements serve solely for the communications purposes. Nevertheless, the transmitter beamformer weights are needed to be computed jointly with each IRS portion. To ease the computational burden of the optimization, \cite{irs_jrc_IRSpartitioning} devises a codebook-based approach, in which the phase shift configuration of the IRS  is selected from a predefined codebook. It is shown that the proposed IRS-assisted DFRC system provides satisfactory spectral efficiency performance on par with that of a system without sensing capabilities.

	Due to the complexity of the joint design, the optimization problem is usually decoupled for DFRC and IRS design. Hence,   alternating optimization techniques are mostly employed (see, e.g., Table~\ref{tableSummary}). Among  these methods, some prioritize the radar~\cite{irs_jrc_txrxbeamforming,irs_jrc_systemsJournal} (communication~\cite{irs_jrc_DiscretePhaseShift,irs_jrc_SecrecyRateOptMishra}) performance with a desired communications (radar) constraint while a joint performance metric optimization is also available~\cite{irs_jrc_interferenceMitigation_Wang,irs_jrc_multipleIRS}.
	
\begin{figure}[t]
		\centering
		{\includegraphics[draft=false,width=.9\columnwidth]{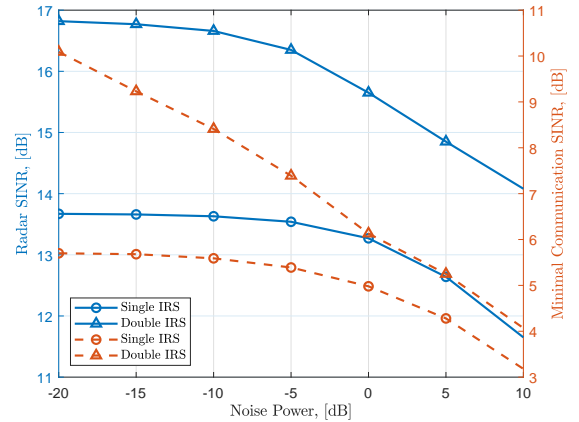} }
		\caption{Radar and communication SINR versus noise power. } \vspace{-10pt}
		\label{figRadarCommSINR}
	\end{figure}

	In \cite{irs_jrc_txrxbeamforming}, alternating optimization of the transmit beamforming and the IRS is performed via semi-definite relaxation (SDR). The detection of the multiple radar targets is prioritized as the beampattern gain of the IRS is maximized with a communications SNR constraint for a single user.  Similar approach is also followed by~\cite{irs_jrc_systemsJournal} for a single user and a target by optimizing the radar SNR with a communications SNR constraint. It is reported that the IRS-assisted DFRC system achieves approximately $10$ dB gain in radar SINR and over non-IRS case. \textcolor{black}{For multi-IRS-assisted ISAC scenario, Fig.~\ref{figRadarCommSINR} shows the radar and minimal communication SINR with respect to the noise power~\cite{irs_jrc_multipleIRS}. It can be seen that the use of double IRS provides about $4$ dB extra gain for both radar and communication SINR.}
	
	%

	The communications performance is prioritized in~\cite{irs_jrc_DiscretePhaseShift}, wherein the interference among the communications users is adopted as a performance metric to design the DFRC and IRS parameters  while constraining the radar performance with the CRLB as the desired target DoA estimation MSE. However, this approach requires the knowledge of the target DoAs for the computation of CRLB. Therefore, the DFRC is first needed to operate in radar mode to obtain the target DoAs, and \cite{irs_jrc_DiscretePhaseShift} employs the well-known \emph{mu}ltiple-\emph{si}gnal \emph{c}lassification  (MUSIC) algorithm for this purpose~\cite{elbir2021JointRadarComm}. In multi-user scenario, \cite{irs_jrc_DiscretePhaseShift} achieves a significant improvement ($\sim80 \%$) in the sum-rate as compared to the non-IRS scenario. Furthermore, it is reported that the communications performance is strongly dependent on the radar constraint (e.g., target detection performance via MUSIC~\cite{irs_jrc_DiscretePhaseShift}).
	
	A practical ISAC deployment model is considered in~\cite{irs_jrc_interferenceMitigation_Wang} and \cite{irs_jrc_multipleIRS} for  a single and multi-IRS cases, respectively. Here, the SINR of both radar and communications are optimized. By leveraging multi-carrier DFRC, the SINR of the overall system is maximized while the trade-off between the radar and communications performance is controlled via a tuning parameter~\cite{elbir2021JointRadarComm,jrc_overview_TCOM}. In~\cite{irs_jrc_interferenceMitigation_Wang}, approximately $80\%$ gain is obtained  in IRS-assisted DFRC in terms of spectral efficiency. Moreover, compared to a single IRS scenario, \cite{irs_jrc_multipleIRS} reported that the use of double IRS improves the radar (communications) SINR approximately $3.3$ ($0.9$) dB for the DFRC operating at $10$ GHz with $32$ subcarriers.



	%

	\subsection{RCC Deployment}
	Performing both tasks with a common IRS requires joint optimization of IRS parameters. In the IRS-enabled coexistence (Fig.~\ref{figIRS_ISAC}), the IRS offers additional degrees of freedom to minimize interference (as mentioned in the IRS-assisted radar section). Particularly, in the presence of a direct link between the entities, the IRS offers additional paths for processing either at the user terminals (UTs) or the BS to mitigate interference~\cite{irs_jrc_DiscretePhaseShift}. In the absence of a direct path, multiple IRSs can enable interference cancellation at the UT or the BS~\cite{irs_jrc_multipleIRS}. \textcolor{black}{Thus, the IRS  inherently establishes a communication (sensing) link for RCC even when there is no LoS path toward the user (target).}  A clear example of RCC is shown in Fig.~\ref{figIRS_ISAC2}, where V2 uses IRS1 and IRS2 individually for communications and radar tasks, respectively. 	Without IRS1 (IRS2), V2 may not communicate with (sense) BS (V1). 	In this scenario, the IRS mounted on the exteriors of buildings may be designated as either radar-only or communications-only. Then, the vehicles may simultaneously connect with these IRSs for either of the two applications. The location information of the IRS for each task is made available at the BS and shared with the vehicles. The IRS parameters are configured relying on the location of vehicles and IRS. In order for multiple vehicles to utilize the same IRS, a wideband scenario may be considered and the IRS parameters for each vehicles are designed in different frequency bands~\cite{irs_jrc_interferenceMitigation_Wang}.

	\subsection{Individual Deployment}
	Besides the aforementioned joint designs, in the simplest case, the IRS may be also deployed for  either communications or radar only (Fig.~\ref{figIRS_ISAC}). 
	While IRS-assisted communications in ISAC yields higher spectral/energy efficiency, the IRS-assisted radar in ISAC improves target detection performance~\cite{irs_jrc_systemsJournal,irs_jrc_DiscretePhaseShift}. Furthermore, the IRS may be used by either of the systems in time-multiplexed manner 
	requiring a trade-off between radar and communications performances depending on the application. \textcolor{black}{For example, the status of the users and targets of being NLoS/LoS and distant/near can be determined in the search phase of the radar and the channel training stage of the communication. Then, IRS is reserved for communications if the users are mostly NLoS and the radar targets are LoS and/or near-range~\cite{irs_radarDetectionBuzziTSP}. 	Conversely, the IRS is slotted for radar tasks if targets are distant/NLoS and the communications users are not.   }
	
	Multiplexed IRS facilitates the RCC systems very well. 	Fig.~\ref{figIRS_ISAC2} shows IRS-assisted RCC in a typical roadside scenario involving vehicles (V), pedestrians (P), IRSs, BS, and a UAV. Here, IRS2 is utilized by V1 for radar-only purposes to sense V2, V3, P1 and P2, while a communication link is established for V1 with P1 and BS.  Similarly, IRS3 at the UAV is used by V4 for communication while its automotive radar senses V3 and V5.  
	
	The switching operation between radar and communications tasks requires a frequent tracking of the user and target locations. Thus, IRS phases and amplitudes are optimized on-the-fly as per the usage. The trade-off between the accuracy/importance of both radar and communications tasks is usually controlled with a tuning parameter in most of the existing IRS-assisted ISAC systems (see, e.g., Table~\ref{tableSummary}). The tuning parameter is usually selected between $0$ (radar-only design) and $1$ (communications-only design)~\cite{elbir2021JointRadarComm} to optimize the balance over  the performance metrics related to both radar and communications, such as the signal-to-interference-plus-noise-ratio (SINR)~\cite{irs_jrc_systemsJournal}, Cramer-Rao lower bound (CRLB) of target direction-of-arrivals (DoAs)~\cite{irs_jrc_DiscretePhaseShift}, IRS beampattern gain~\cite{irs_jrc_txrxbeamforming}, MSE of the communications symbols~\cite{irs_jrc_DiscretePhaseShift,irs_jrc_multipleIRS} and the MSE of target direction vector~\cite{irs_jrc_IRSpartitioning}. The optimization process is rather straightforward in narrowband 
	communications~\cite{irs_Overview_Pan2022Aug} and radar~\cite{irs_radarDetectionBuzziTSP}. However, at mm-Wave and higher, wideband processing requires \textcolor{black}{the transmitted waveform} to be optimized jointly for multiple carriers~\cite{jrc_overview_TCOM,irs_jrc_interferenceMitigation_Wang,irs_jrc_multipleIRS}.

	\section{Design Challenges}
	\textcolor{black}{Compared to ISAC-only systems, the integration of IRS and ISAC introduces the specific major signal processing challenges including channel estimation of additional paths from the IRS, handling clutter via IRS-assisted links, incorporating multipath signals through IRS, waveform design and resource allocation by taking into consideration of IRS deployment, and computational complexity of the joint design problem involving IRS parameters. }


	\subsection{Environment Knowledge}
	\textcolor{black}{Prior information of environment is helpful for accurate sensing performance to mitigate the target-related uncertainty.	For instance, the IRS location should be available at the DFRC transmitter so that it generates accurate beams toward IRS~\cite{irs_radarDetectionBuzziTSP,irs_Overview_Pan2022Aug}. While this is straightforward when both DFRC and IRS are stationary, it is  challenging for a mobile scenario. Such applications include DFRC-equipped vehicles and IRS-mounted UAVs (see Fig.~4). The DFRC utilizes the geometry of the scene to control the IRS and decide if the received signals are NLoS/LoS~\cite{irs_radar_activePassiveBeamforming}. In addition, the target location information needs to be acquired before tracking operation.  Thus, the DFRC should first employ DF algorithms (e.g., MUSIC) to estimate the target directions~\cite{elbir2021JointRadarComm}. }
	
	\subsection{Channel Acquisition}
	In communications,	the channel state information (CSI) is estimated to optimize the wireless data transmission. Compared to the conventional massive MIMO networks, IRS-assisted architectures require estimation of multiple channel links, such as BS-IRS and IRS-user. Channel estimation for IRS-assisted ISAC is even more challenging because of joint processing of user messages with radar backscatter. Further, in dynamic indoor or vehicular scenarios, channel estimation and tracking problems are more exacerbated. In addition, while a MIMO radar typically emits omnidirectional probing signals to search for possible targets, pilot signals are transmitted to users to estimate CSI. To resolve this, radar probing signals may be used as the pilot signals. In this way, the pilot signals are received by the users via IRS and the estimated CSI is fedback to the BS~\cite{irs_Overview_Pan2022Aug}. \textcolor{black}{The number of pilot signals increases proportionally with the number of IRS elements and the number of antennas at the BS and the user. In such a scenario, partitioning the antenna/IRS elements as group-of-subarrays (GoSA) yields low-cost and low-complexity system, especially for massive/ultra-massive MIMO applications~\cite{elbir2021JointRadarComm}.}  \textcolor{black}{Nevertheless, an important ISAC feature is a common waveform for both sensing and communications that takes into account the interference and clutter. In this context, imparting the IRS technologies to the ISAC brings more challenges to the waveform design problem. For instance,	in~\cite{jrc_irs_waveform_Liu2022May}, joint waveform and beamformers are studied for IRS-assisted ISAC, wherein space-time adaptive processing (STAP) technique yields additional DoF and adaptive clutter suppression.}

	\begin{figure}[t]
		\centering
		{\includegraphics[draft=false,width=.9\columnwidth]{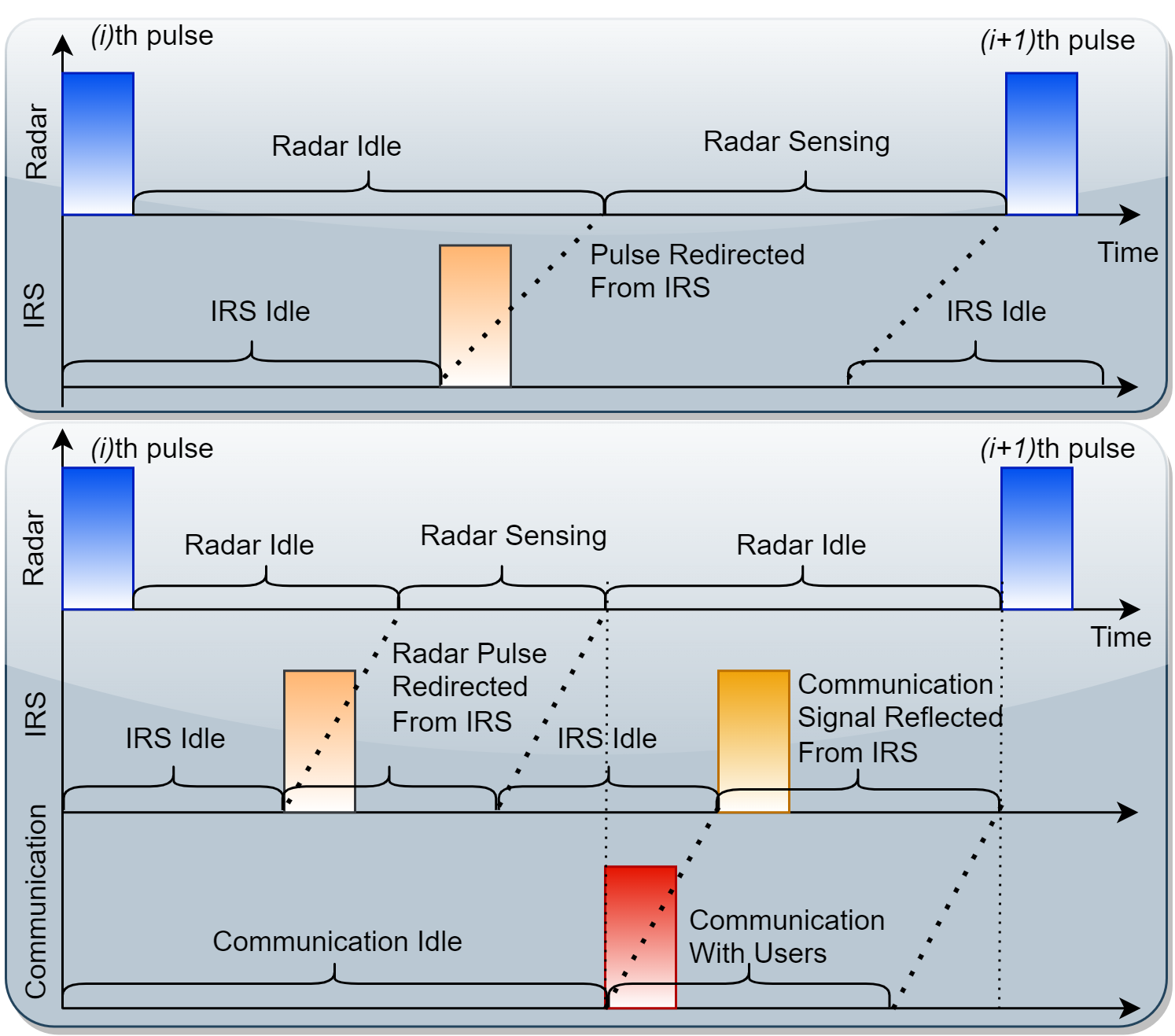} }
		\caption{Radar pulse transmission and communications in time-division manner for IRS-assisted  radar (top) and IRS-assisted ISAC (bottom). } \vspace{-10pt}
		\label{figPulseTransmission}
	\end{figure}

	\subsection{Waveform Design and Resource Management}
	\textcolor{black}{In ISAC, waveform design and resource management are one of the most challenging signal processing issues since the designed should take into account the integration of both sensing and communication functionalities.  In the absence of IRS, the resource allocation in DFRC design has been performed in different transmission schemes, e.g., non-overlapping (time/frequency/spatial/code-division multiplexing) or fully unified (sensing-centric, communication-centric or joint design) architectures~\cite{jrc_overview_TCOM}. While the former schemes are easier to implement, they yield low system efficiency; the latter techniques have guaranteed sensing/communications performance at the cost of high signal processing and hardware complexity. Nevertheless, the main goal of ISAC is to achieve a unified waveform design to perform both sensing and communication task as well as taking into account the interference and clutter. In this regard, imparting the IRS technologies to the ISAC makes the waveform design even more challenging.	In~\cite{jrc_irs_waveform_Liu2022May}, joint waveform and beamforming optimization is studied for IRS-assisted ISAC, wherein the STAP technique is employed to achieve additional DoF and adaptive clutter suppression.}	In Fig.~\ref{figPulseTransmission} a simple radar pulse transmission and communications method is illustrated in time-division manner for both IRS-assisted radar and IRS-assisted ISAC. Here, the radar first emits a pulse which is reflected from the IRS toward the targets, then listens the incoming signals. Then, the communications signal is transmitted, which is reflected via IRS toward the communications users. \textcolor{black}{Despite the simplicity of this time-division scheduling, it does not fully leverage the integration of sensing and communication, which requires further research in waveform design for IRS-assisted ISAC.  }

	\subsection{Clutter Suppression and Multi-User Interference}
	While the communications scenario suffers from multi-user interference (MUI), the radar backscatter often entails dealing with clutter or reflections from unwanted targets, e.g., buildings, ground, and vegetation. \textcolor{black}{In a cluttered environment, the IRS is helpful by creating an additional LoS path to improve the performance against target blockage, especially in indoor sensing applications~\cite{irs_jrc_multipleIRS,irs_radar_activePassiveBeamforming}.  By exploiting the signal-dependence of the clutter interference, the effect of clutter can be suppressed by utilizing the clutter covariance matrix, which requires the prior knowledge of the clutter. In addition, the BS can transmit a dedicated sensing signal in order to effectively suppress the clutter in IRS-assisted ISAC scenario~\cite{irs_jrc_clutter_Liao2022Aug}. Furthermore, STAP technique is used in~\cite{jrc_irs_waveform_Liu2022May} for clutter suppression by exploiting additional DoFs in both spatial and temporal domain. }
	
	\textcolor{black}{Another challenging scenario may be prioritization of the targets which are detected/tracked via the same IRS. This problem may require the prior information about the radar scene for the DFRC to prioritize the targets and schedule the radar resources accordingly.       }

	\subsection{Security}
	In most of the DFRC applications, the radar and communications signals are transmitted for both services using the same frequency band~\cite{jrc_overview_TCOM}. Thus, the delivery of information to both communications users and radar targets becomes more challenging in ISAC with the increasing the chance of security violation by the unauthorized users or eavesdroppers. \textcolor{black}{Thanks to its capability of configuring the wireless channels and increased DoF, the usage of IRS can significantly enhance the physical layer security by controlling the IRS elements to direct the signal toward legitimate users meanwhile degrade the reception at the eavesdroppers.
		Hence, the rate of reliable information delivered to the intended users/targets is maximized while the eavesdroppers are kept as ignorant as possible. One possible way of achieving this is to employ artificial noise to minimize the SINR at the eavesdroppers while satisfying an SINR threshold at the legitimate users/targets. In the initial work in~\cite{irs_jrc_SecrecyRateOptMishra},  a simultaneous secure and communication is considered in a IRS-assisted DFRC scenario; it was shown that DFRC systems enjoy a significant improvement in terms of secrecy rate with the aid of IRS.}

	\section{Summary and Future Outlook}
	We provided a synopsis of recent developments toward enabling IRS-assisted radar and ISAC technologies. The IRS deployment provides significant improvements in terms of target detection, interference suppression, spectral efficiency and security. 	
	In particular, IRS-aided radar improves NLoS target detection, target parameter estimation and interference suppression. This is particularly helpful in automotive radar applications, wherein targets may fall into the shadow range of the radar in densely populated urban environments. The IRS may also boost the received signal power at the vehicle, thereby, improving the target detection. The IRS-reflected signals could also be utilized to suppress the interference from unauthorized users/targets. 
	
	These advantages notwithstanding, use of IRS introduces a more complex system architecture that necessitates advanced signal processing techniques for joint DFRC and IRS design.  From an algorithmic perspective, joint optimization of DFRC and IRS parameters requires novel and computationally-efficient signal processing techniques, and the fundamental limits of radar and communication performance need to be assessed for the plethora of configurations. 
	From an operational perspective,  IRS-aided DFRC assumes knowledge of IRS location to control/design the IRS parameters. While this information is accurately acquired in static deployment, the same is challenging for mobile architectures, e.g., DFRC/IRS mounted on vehicles/UAVs. Finally, from a emerging scenario perspective, possible deployments in mm-Wave and THz frequencies invariably entail massive number of active (DFRC) and passive (IRS) elements. In this respect, the subarrayed architectures and data-driven techniques, e.g., deep learning (DL), can be very helpful. The DL-based techniques can also be effective for blind channel estimation without feedback and reduce the number of pilot signals particularly for hardened channels with large IRSs. Each of the aforementioned scenarios open new research directions bringing its fair share of algorithmic and practical nuances.



	\balance

	%
	%

	
	\bibliographystyle{IEEEtran}
	\bibliography{references_090}

	\begin{IEEEbiographynophoto} {Ahmet M. Elbir} (Senior Member, IEEE)  is currently a Research Fellow at 	the Interdisciplinary Centre for Security, Reliability and Trust (SnT), University of Luxembourg; and		Duzce University, Turkey.
	\end{IEEEbiographynophoto}
	
	\begin{IEEEbiographynophoto} {Kumar Vijay Mishra} (Senior Member, IEEE)  is currently,  a National Academies Harry Diamond Distinguished Fellow at the U. S. Army Research Laboratory.
	\end{IEEEbiographynophoto}

	\begin{IEEEbiographynophoto} {M. R. Bhavani Shankar} (Senior Member, IEEE) is currently an Assistant Professor leading the Signal Processing Applications in Radar and Communications (SPARC) in the SnT, University of Luxembourg.
	\end{IEEEbiographynophoto}
	
	\begin{IEEEbiographynophoto} {Symeon Chatzinotas} (Fellow, IEEE) is currently the Head of the research group SIGCOM at SnT, University of Luxembourg.
	\end{IEEEbiographynophoto}
	

	%

\end{document}